\documentclass[12pt]{article}
\usepackage{a4wide}
\usepackage{amssymb,amsfonts}
\usepackage{graphics,psboxit,amsmath}
\usepackage{subfigure}
\usepackage{graphicx}
\usepackage{verbatim}

\makeatletter
\@addtoreset{equation}{section}
\makeatother

\def\be{\begin{equation}}
\def\ee{\end{equation}}
\def\ba{\begin{eqnarray}}
\def\ea{\end{eqnarray}}

\def\del{\partial}

\def\k{\kappa}
\def\r{\rho}
\def\a{\alpha}

\def\b{\beta}

\def\g{\gamma}

\def\d{\delta}

\def\th{\theta}

\def\m{\mu}
\def\n{\nu}
\def\om{\omega}
\def\Om{\Omega}

\def\L{\Lambda}
\def\s{\sigma}

\def\cN{{\cal N}}

\def\no{\noindent}

\def\qq{\qquad}

\def\IR{\relax{\rm I\kern-.18em R}}

\def\inv{^{\raise.0ex\hbox{${\scriptscriptstyle -}$}\kern-.05em 1}}

\def \ha {{\frac{1}{2}}}

\def \ov {\over}

\newcommand{\tr}{\mbox{tr}}

\begin{document}

\begin{flushright}
June 2011 \hfill
FPAUO-11/02 \\
CPHT-RR011.0411
\end{flushright}

\vspace{1.5truecm}

\centerline{\LARGE \bf ABJM Baryon Stability and Myers effect}
\vspace{1.3truecm}

\centerline{
    {\large \bf Yolanda Lozano${}^{a,}$}\footnote{
                                   {\tt ylozano@uniovi.es}},
  \ {\large \bf Marco Picos${}^{a,}$}\footnote{
    \tt picosmarcos@uniovi.es                                  }}

\vspace*{0.5cm}
\centerline{
 {\large \bf Konstadinos Sfetsos${}^{b,}$}\footnote{
                                  {\tt sfetsos@upatras.gr}}
                                \  and\
    {\large \bf Konstadinos Siampos${}^{c,}$}\footnote{
                                   {\tt ksiampos@cpht.polytechnique.fr}}
    }

\vspace{.7cm}

\centerline{{\it ${}^a$Department of Physics,  University of Oviedo,}}
\centerline{{\it Avda.~Calvo Sotelo 18, 33007 Oviedo, Spain}}

\vspace{.4cm}
\centerline{{\it ${}^b$Department of Engineering Sciences, University of Patras,}}
\centerline{{\it 26110 Patras, Greece}}

\vspace{.4cm}
\centerline{{\it ${}^c$Centre de Physique Th\'eorique, Ecole Polytechnique, CNRS }}
\centerline{{\it 91128 Palaiseau, France}}

\vspace{1truecm}

\centerline{\bf ABSTRACT}
\vspace{.5truecm}

\noindent
We consider magnetically charged baryon vertex like configurations in $AdS_4\times CP^3$
with a reduced number of quarks $l$.
 We show that these configurations are solutions to the classical equations of motion
and are stable beyond a critical value of $l$.
 Given that the magnetic flux dissolves D0-brane charge it is possible to give
a microscopical description in terms of D0-branes expanding
into fuzzy $CP^n$ spaces by Myers dielectric effect. Using this description
we are able to explore the region of finite 't Hooft coupling.\\
\\

\newpage

\tableofcontents

\def\baselinestretch{1.2}
\baselineskip 20 pt

\section{Introduction}

The $AdS_4/CFT_3$ duality relates the Type IIA superstring on $AdS_4\times CP^3$ to an ${\cal N}=6$ quiver
Chern-Simons-matter theory with gauge group $U(N)_k \times U(N)_{-k}$ known as the ABJM model \cite{ABJM}.
Like its $AdS_5/CFT_4$ counterpart it is a strong/weak coupling duality, with 't Hooft coupling $\lambda=N/k$.
 Being the superpotential coupling proportional to $k^{-2}$, an appropriate large $k$ limit $N\ll k^5$
allows for a weak coupling regime. The Type IIA theory is then  weakly curved when $k\ll N$.

The $CP^3$ space has $H^q(CP^3)=\mathbb{R}$ for even $q$. Therefore it is possible to have
D2, D4 and D6 particle-like branes wrapping a topologically non-trivial cycle.
 In $AdS_4\times CP^3$ these branes were already discussed in \cite{ABJM},
and their interpretation in the context of the CFT dual given.
The D6-brane wrapped on the entire $CP^3$ is the analogous of the baryon vertex
in $AdS_5\times S^5$ discussed by Witten in \cite{Witten}.
Due to the $F_6$ flux of the background it has a tadpole that has to be
cancelled with $N$ fundamental strings ending on it, which correspond to $N$ external quarks
on the boundary of $AdS_4$. Similarly, the D2-brane wrapped on a $CP^1\subset CP^3$
captures the $F_2$ flux of the $AdS_4\times CP^3$ background,
and develops a tadpole that has to be cancelled with $k$ fundamental strings ending on it.
The field theory interpretation of this brane is as a 't Hooft monopole,
realized as a ${\rm Sym}_k$ product of Wilson lines.
The D4-brane wrapped on a $CP^2\subset CP^3$ does not capture any
of the background fluxes, and it is gauge invariant. It is dual to the
di-baryon operator \cite{GK,BHK}, which has the same baryon charge
and dimension to agree with the gravity result.

These gravitational configurations admit a natural generalization
by allowing non-trivial worldvolume gauge fluxes \cite{GLR}.
These generalizations have been proposed as candidates
for holographic anyons \cite{Hartnoll} in ABJM \cite{KL}, and are therefore
of potential interest for AdS/CMT applications.
Allowing for a non-trivial worldvolume magnetic flux has the
effect of adding lower dimensional brane charges to the configurations,
in particular D0-brane charge. This modifies how the brane captures the
background fluxes in a way that depends on the induced charges, such that,
in some cases, additional fundamental strings are required to cancel the
 worldvolume tadpoles. The D2 and D6-branes are only stable if the induced charges lie below some upper bound. In turn, 
the D4-brane with flux behaves quite differently from the zero charge case, since it now requires fundamental strings 
ending on it. Given that in the presence of a non-trivial magnetic flux all these branes require fundamental strings 
ending on them we will loosely refer to them as baryon vertex like configurations.

In this paper we further generalize these constructions by reducing the
 number of strings that stretch between the brane and the boundary
of $AdS_4$, i.e. the number of quarks. It was shown in \cite{BISY,Imamura}
that in $AdS_5\times S^5$ perfect baryon vertex classical solutions
to the equations of motion exist for a number of quarks $l$ satisfying $5N/8 \leqslant l \leqslant N$.
Although one would expect that bound states of quarks should be singlets
of the gauge group the analysis of the stability against fluctuations
confirms that the configurations are stable for a number of
quarks $0.813 N \leqslant l \leqslant N$ \cite{SS}. It is likely that this will
not be the case in other theories with reduced supersymmetry.

It is one aim of this paper to perform a similar analysis
for magnetically charged baryon vertex like configurations with reduced number
of quarks in $AdS_4 \times CP^3$. Our analysis will reveal that also in this
case baryon vertex like classical solutions exist that are moreover stable against fluctuations.

In order to be able to use the probe brane approximation in the study of the
dynamics we will consider a uniform distribution of strings on a $CP^{\frac{p}{2}}$
 geometrical shell, with $p=2,4,6$. This will be our particular profile for the distribution
of quarks inside the baryon vertex configuration. Although this choice completely breaks
supersymmetry we will be able to ignore the strings backreaction \cite{Imamura,Imamura2,CGS}.

The fact that the magnetized branes have dissolved
D0-branes in their worldvolumes hints at the existence of a microscopical
description in terms of non-Abelian $n$ D0-branes polarizing due to Myers
dielectric effect \cite{Myers}. This description allows to explore the
 configurations in the region where $N\ll n^{\frac{4}{p}}\, k$, and is therefore complementary
to the supergravity description in terms of probe branes. We will see that
classical stable solutions still exist in this regime. Moreover, we will show that the flat half-integer
 $B_2$ field that is required by the Freed-Witten anomaly in the di-baryon \cite{AHHO} has to be introduced 
already at the classical level so that a $CP^2$ non-spin manifold can be recovered in the large $n$ limit.

The organization of the paper is as follows:
We start in section 2 by summarizing some of the properties of the
magnetized baryon vertex like configurations constructed in \cite{GLR}.
In section 3 we reduce the number of quarks and find the values for which
classical configurations still exist.  In section 4 we perform the
stability analysis under small fluctuations. Section 5 is devoted to the
microscopical description. This description will confirm the existence of non-singlet
classical stable solutions when $N\ll k^5$.  An interesting output of this
analysis will be the derivation of new higher curvature dielectric couplings
not predicted before in the literature.
Finally, in section 6 we summarize our results and discuss further directions.
We have written appendix A, containing a number of useful results on
the $AdS_4 \times CP^3$ background and also appendix B with the
computation of the K\"ahler form for the fuzzy $CP^{\frac{p}{2}}$, used in the main text.

\section{Magnetically charged baryon vertex configurations in  $AdS_4\times CP^3$ spaces}

It was shown in \cite{GLR} that it is possible to construct more general monopole,
di-baryon and baryon vertex configurations in $AdS_4\times CP^3$ if the particle
like branes carry lower dimensional brane charges induced by a non-trivial
magnetic flux $F={\cal N}J$, where $J$ is the K\"ahler form of the $CP^3$.
For the D2 and D6 branes the effect of the magnetic flux is to allow
the construction of similar monopole and baryon vertex configurations
with D0-brane charge and a different number of fundamental strings attached.
 Indeed the study of the dynamics reveals that the
configurations are stable if the
 magnetic flux does not exceed some maximum value, for which the
configurations reduce to radial fundamental strings (free quarks) plus the wrapped D-brane.

The di-baryon is more substantially modified by the presence of the magnetic flux,
capturing the $F_2$ flux and developing a tadpole. In this case the study of the
dynamics shows that the D4-brane with the fundamental strings attached is stable
if the magnetic flux takes values in a given interval, at the limits of which
the configuration ceases to be stable and reduces to free quarks plus the D4-brane.
 This is consistent with the fact that the D4-brane with F-strings does not exist
for zero magnetic flux. Moreover, since the D4-brane wraps a non-spin manifold
it must carry a half-integer worldvolume magnetic flux due to the Freed-Witten
anomaly \cite{FW}. In order to still keep its dual interpretation as a di-baryon
it was proposed in \cite{AHHO} that a flat half-integer $B_2$-field should
be switched on in the dual background in order to cancel the contribution of the 
Freed-Witten worldvolume magnetic flux. 

A question that remained open after the study in \cite{GLR} was the interpretation of the magnetized D$p$-branes in the field theory. A difficulty comes from the expected lack of SUSY for the D2 and D6-branes. In turn the D4-brane with flux forms a threshold BPS intersection with the D0-branes. Therefore one could expect that a supersymmetric spiky solution exists and one could give an interpretation to the bounds in the gauge theory dual. As shown in \cite{GLR} the maximum (and minimum, if applicable) values of the magnetic flux are functions of $\sqrt{\lambda}$, with $\lambda$ the 't Hooft coupling, for all branes. This suggests an origin on the conformal symmetry of the gauge theory. Ultimately one would expect a connection between the existence of these bounds and the stringy exclusion principle of \cite{MS}.
 
We summarize next the energies and charges carried by the various branes. In
order to set up the notation
a short review of the $AdS_4\times CP^3$ background is given in appendix A. We will use Poincar\'e coordinates to parameterize $AdS_4$ throughout the paper.

\subsection{Charges and energies}

The computation of the energy of a D$p$-brane
in $AdS_4\times CP^3$ wrapped on a $CP^{\frac{p}{2}}$ cycle of
the $CP^3$ with $p=2,4$ and $6$, in the presence of a magnetic flux $F={\cal N}J$,
with ${\cal N}\in 2\mathbb{Z}$, was done in \cite{GLR}.
We review this result and show that the equations of motion are satisfied for $F={\cal N}J$.

The DBI action is
\be
S_p=-{T_p}\int d^{p+1}\xi\,  e^{-\phi}  \sqrt{|\det(P[g+2\pi {\cal F}])|}\ ,\qquad T_p={1\ov (2\pi)^p}\ ,
\ee
where ${\cal F}=F+\frac{1}{2\pi}B_2$ and we set $\ell_s=1$.
The equations of motion arising from varying the gauge potential are given by
\be
\del_\a\left(\sqrt{|\det P([g+2 \pi {\cal F}])|}\ (P[g+2 \pi {\cal F}])^{-1[\a\b]}\right)=0\ ,
\label{fjh22}
\ee
where $[\a\b]$ denotes the antisymmetric part.
Identifying the world-volume coordinates with the angles of the
various $CP$-cycles as indicated in
appendix A and considering static solutions independent of the $\xi^i$'s we find an
induced metric 
\be
ds^2_{\rm ind}=- \frac{16{\rho^2}}{L^2} d\tau^2 + L^2 ds^2_{CP^{\frac{p}{2}}}\ .
\ee
Using that in our case ${\cal F}$ is proportional to the K\"ahler form, since $F={\cal N}J$ and $B_2=-2\pi J$, we can easily
 prove that the equations of motion are satisfied. 
If $M$ is an antisymmetric $p\times p$ matrix satisfying $M^2=-c\ \mathbb{I}$, where for consistency 
$\displaystyle c=-{1\ov p} {{\rm Tr}( M^2)}$, one can show that
$(\mathbb{I}+M)^{-1}={\mathbb{I}-M\ov 1+c}$, and, moreover, due to the fact that $M$ is antisymmetric: 
$\det(\mathbb{I}+M)=(1+c)^{\frac{p}{2}}$. Using these identities we find that
 $\partial_\a(\sqrt{g}J^{\a\b})=0$, where $g$ is the metric
on $CP^{\frac{p}{2}}$, or equivalently $\nabla_\a J^{\a\b}=0$. The latter is
the condition for having a K\"ahler manifold and therefore it is automatically satisfied. Also we find
that the DBI action is given by (we use $c=(2 \pi \cN)^2$)
\begin{equation}
\label{energymac}
S_{DBI}^{Dp} = -\frac{T_p}{g_{s}}\int d^{p+1}\xi \sqrt{-{\rm det}(g+2\pi {\cal F})}=- Q_p\int d\tau {2\r\ov L}\ ,
\end{equation}
where
\be
\label{Qp1}
Q_p=\frac{T_p}{ g_s}\ {\rm Vol}(CP^{\frac{p}{2}}) \ \left(L^4 + (2\pi)^2({\cal N}-1)^2\right)^{\frac{p}{4}}\, , \qquad {\rm for} \quad p=2,6
\ee
and 
\be
\label{Qp2}
Q_4=\frac{T_4}{ g_s}\ {\rm Vol}(CP^{2}) \ \left(L^4 + (2\pi{\cal N})^2\right)\, ,
\ee
since in this case $B_2$ cancels the contribution of the Freed-Witten vector field, such that
${\cal F}=F_{FW}+{\cal N}J+\frac{1}{2\pi}B_2={\cal N}J$. Also, in this case
\begin{equation}
S_{DBI}^{D4}=-\frac{T_4}{g_{s}}\int d^5\xi \sqrt{-{\rm det}(g+2\pi {\cal F})}=
-\frac{T_4}{g_{s}}\int d^5 \xi \, \sqrt{|g_{tt}|} \sqrt{g_{\mathbb{P}^2}}\, \Bigl(L^4+2(2\pi)^2{\cal F}_{\alpha\beta}{\cal F}^{\alpha\beta}\Bigr)
\end{equation}
The volume of the $CP^{\frac{p}{2}}$ is given by
\be
{\rm Vol}(CP^{\frac{p}{2}})={\pi^{\frac{p}{2}}\ov \left(\frac{p}{2}\right)!}\ .
\ee
From (\ref{Qp1}) and (\ref{Qp2}) it is clear that $\cN^2$ is comparable to $L^4\gg 1$.

Analyzing the Chern-Simons actions one can also show that the magnetic flux has the effect of dissolving lower dimensional brane
 charge in the D$p$-branes. For instance the D4-brane has D2 and D0-brane charges dissolved, as can be seen from the couplings:
 \begin{equation}
S_{CS}^{D4}=2\pi\, T_4 \int_{\mathbb{R}\times \mathbb{P}^2} C_3\wedge F=\frac{{\cal N}}{2} \, T_2 \int C_3
\end{equation}
and
\begin{equation}
S_{CS}^{D4}=\frac12 (2\pi)^2\, T_4 \int_{\mathbb{R}\times \mathbb{P}^2}  
C_1 \wedge F\wedge F= \frac{{\cal N}^2}{8}\, T_0 \int_{\mathbb{R}} C_1\, ,
\end{equation}
respectively. In general the number of D$s$-branes dissolved in the worldvolume of a D$p$ is given by
\cite{GLR}
\begin{equation}
\label{nDqs}
n=\frac{{\cal N}^{\frac{p-s}{2}}}{2^{\frac{p-s}{2}}(\frac{p-s}{2})!}\, .
\end{equation}
Both the D4 and D6-branes have $CP^1$ D2-branes dissolved. Therefore in the presence of a magnetic flux they
capture the $F_2$ flux and develop a tadpole with charge
\begin{equation}
\label{kcharge}
q=k\, \frac{{\cal N}^{\frac{p}{2}-1}}{2^{\frac{p}{2}-1}(\frac{p}{2}-1)!}
\end{equation}
More explicitly, for the D4-brane we have that 
\begin{eqnarray}
S_{CS}^{D4}&=& \frac12 (2\pi)^2\, T_4 \int_{\mathbb{R}\times \mathbb{P}^2} P[F_2]\wedge F \wedge A=
2\,(2\pi)^2 T_4 \, k\,{\cal N}\int_{\mathbb{R}\times \mathbb{P}^2} J\wedge J\wedge A\nonumber\\
&=&k\,\frac{{\cal N}}{2}\, T_{F1} \int dt A_t
\end{eqnarray} 
The analogous coupling for the D6-brane is
\begin{equation}
S_{CS}^{D6}=\frac16 (2\pi)^3\, T_6 \int_{\mathbb{R}\times \mathbb{P}^3} P[F_2]\wedge F \wedge F \wedge A=
k\, \frac{{\cal N}^2}{8}\, T_{F1} \int dt A_t\, .
\end{equation} 
Note however that for the D6-brane the couplings 
$\int_{D6} F_2\wedge B_2\wedge B_2\wedge A$ and
$\int_{D6} F_2\wedge F\wedge B_2\wedge A$ in its CS action contribute as well to its 
$k$ charge. In the absence of magnetic
flux it was shown in \cite{AHHO} that the contribution from $\int_{D6} F_2\wedge B_2\wedge B_2\wedge A$
 is
cancelled from the higher curvature coupling  \cite{GHM,CY,BBG}
\begin{equation}
\label{Higher_Curv}
S_{h.c.}^{D6}=\frac32  (2\pi)^5\, T_6 \int C_1\wedge F \wedge
 \sqrt{\frac{\hat{\mathcal{A}}(T)}{\hat{\mathcal{A}}(N)}}\, ,
\end{equation}
where $\hat{\mathcal{A}}$ is the $A$-roof (Dirac) genus
\begin{equation}
\hat{\mathcal{A}}=1-\frac{\hat{p}_1}{24}+\frac{7\, \hat{p}_1^2-4\,\hat{p}_2}{5760}+\cdots
\end{equation}
and the Pontryagin classes are written in terms of the curvature of the corresponding bundle as
\begin{equation}
\hat{p}_1=-\frac{1}{8\pi^2}\, {\rm Tr}\, R^2\, , \qquad \hat{p}_2=\frac{1}{256\,
\pi^4}\,\Big(({\rm Tr}\, R^2)^2-2\, {\rm Tr}\, R^4\Big)\, .
\end{equation}
This charge cancellation is consistent with the dual interpretation of the D6-brane as a baryon vertex.
For a non-vanishing magnetic flux the term $\int_{D6} F_2\wedge F\wedge B_2\wedge A$
 contributes however with $-k{\cal N}/4$ units of F-string charge, as shown in \cite{GLR}.
Therefore, adding the $N$ units induced by the $F_6$ flux, 
\begin{equation}
\label{F1_no_flux}
S_{CS}^{D6}=2\pi \, T_6\int_{R\times \mathbb{P}^3}P[F_6]\wedge A= N\, T_{F1}\int dt A_t\, ,
\end{equation}
not captured by the
other branes, we find that the total F-string charge carried by the D6-brane is given by
\begin{equation}
\label{qD6}
q_{D6}=N + k\, \frac{{\cal N}({\cal N}-2)}{8}
\end{equation}
Note that this is always an integer due to the quantization condition
\begin{equation}
\frac{1}{2\pi}\int F=\frac{{\cal N}}{2}\in \mathbb{Z}
\end{equation}

\section{Varying the number of fundamental strings}

It was shown in \cite{BISY,Imamura} that the baryon vertex in
$AdS_5\times S^5$ can be generalized such that the number of quarks $l$ lies
in the interval $5N/8\leqslant l\leqslant N$. These configurations are not only perfect classical
 solutions to the equations of motion but for $0.813\,N\leqslant l\leqslant N$
are stable against fluctuations \cite{SS}.
In this section we generalize the construction in \cite{BISY,Imamura}
to the baryon vertex like configurations discussed in the previous section.
We will see that in all cases there exist configurations with a reduced
number of quarks that are solutions to the classical equations of motion.

We consider a classical configuration consisting on a D$p$-brane wrapped
on $CP^{\frac{p}{2}}$, located at $\rho=\rho_0$, $l$ strings stretching from $\rho_0$
to the boundary of $AdS_4$ and $(q-l)$ straight strings that go from $\rho_0$
to 0. The configuration is depicted in Figure 1.
Further, we switch on the magnetic flux $F={\cal N}J$, with $J$ the K\"ahler form of the $CP^3$.
Taking the gauge  $\tau=t$, $\sigma=\rho$ for the worldsheet coordinates of the
string, 
the Nambu-Goto action of the $l$ fundamental strings is given by \cite{Maldacena}
\begin{equation}
S_{l F1}=-l\, T_{F1}\int dt d\rho \, \sqrt{1+\frac{16\rho^4}{L^4}{r^\prime}^2}
\end{equation}
where $r$ is the radius of the configuration at the boundary of $AdS_4$.
The equations of motion then reduce to
\begin{equation}
\label{eqmotion2}
\frac{16\rho^4 r^\prime}{L^4\sqrt{1+\frac{16\rho^4}{L^4}{r^\prime}^2}}=c=\frac{4\rho_1^2}{L^2}
\end{equation}
where the constant has been fixed demanding that $r^\prime=\infty$ at the
 turning point of each string, $\rho_1$. The turning point is such that
 $0\leqslant \rho_1\leqslant \rho_0$. From (\ref{eqmotion2})
\begin{equation}
\label{eqmotion3}
r^\prime=\frac{L^2\rho_1^2}{4\rho^2\sqrt{\rho^4-\rho_1^4}}\equiv r^\prime_{\rm cl}
\end{equation}

\begin{figure}[!t]
\begin{center}
\begin{tabular}{cc}
\includegraphics[scale=.6]{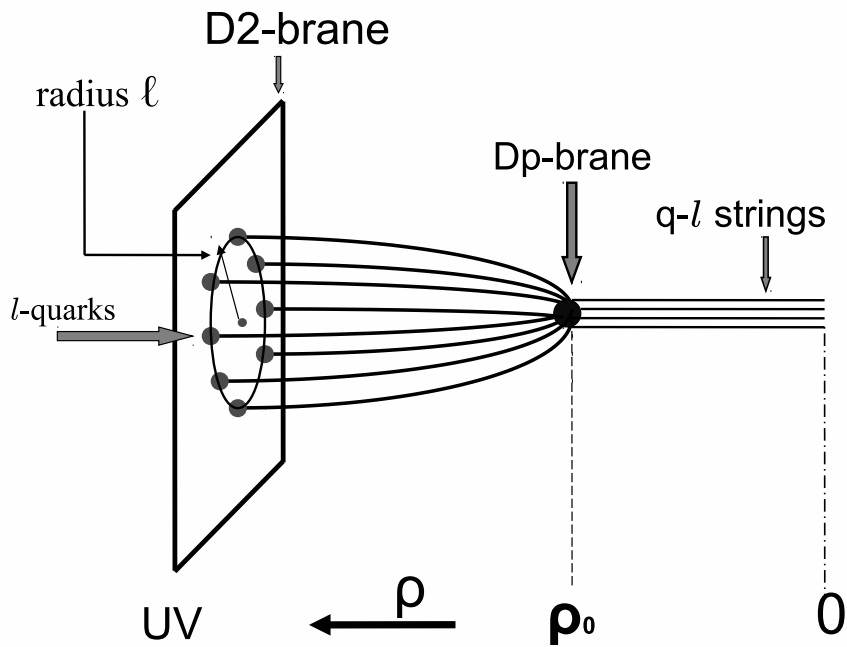}
\end{tabular}
\end{center}
\vskip -2.5 cm \caption{A baryon configuration with $l$-external quarks placed on a
circle of radius $\ell$
at the boundary of AdS space, each connected to a $Dp$-brane wrapped on a
$CP^{\frac{p}{2}}$ located at $\rho=\rho_0$, and $q-l$ straight
strings ending at $0$.}
\end{figure}

\noindent Defining $a\equiv \frac{l}{q}$,
the boundary equation reads
\begin{equation}
\label{boundary2}
\frac{1}{a}\sqrt{1-\beta^2}+\frac{1-a}{a}=\sqrt{1-\frac{\rho_1^4}{\rho_0^4}}
\end{equation}
where we defined \cite{GLR}
\begin{equation}
\label{beta}
\sqrt{1-\beta^2}\equiv\frac{2 Q_p}{L\,q\,T_{F_1}}\, .
\end{equation}
We then must have
\begin{equation}
\label{bound}
\frac{2 Q_p}{L\,q\,T_{F_1}}\leqslant 1
\end{equation}
in order to find a stable configuration.
Since $Q_p$ (and also $q$, for the D4 and D6-branes), are functions
 of ${\cal N}$ this condition imposes a bound on the magnetic flux that can
 be dissolved on the worldvolume. For the D2 and D6-branes ${\cal N}$ must
 lie below some upper bound, for which $\beta=0$. For the D4 the magnetic
flux must also lie above a lower bound, for which $\beta=0$ as well. This
is consistent with the fact that the D4-brane with fundamental strings
attached only exists for non-zero magnetic flux.

For the values of the magnetic flux allowed by equation (\ref{bound}) we must
 still fulfill the boundary equation (\ref{boundary2}), and this implies that
\begin{equation}
\label{bound2}
q\sqrt{1-\beta^2}+q-l\leqslant l\qquad \Leftrightarrow \qquad l\geqslant \frac{q}{2}(1+\sqrt{1-\beta^2})=l_{\rm min}
\end{equation}
This condition determines the minimum value of strings that can form the
baryon vertex like configuration. Note that $l_{\rm min}$ is a function of the
 magnetic flux, and is such that it decreases with $\beta$. For the D2 and D6-branes $\beta$
is maximum for zero magnetic flux, for which $l_{\rm min}$ reaches its minimum value:
 $l_{\rm min}=\frac{q}{2}(1+\frac{1}{2\pi})$, $l_{\rm min}=\frac{q}{2}(1+\frac{1}{6\pi})$,
respectively. Recall that for this value of the magnetic flux the configuration
is maximally stable \cite{GLR}. For the D4-brane $\beta$ is maximum
when $\frac{{\cal N}}{L^2}=\frac{1}{2\pi}$, which also corresponds to the
 most stable configuration. For this value of the magnetic flux
$l/q$ is minimum\footnote{Recall that in this case $q=k{\cal N}/2$.}, and
 one finds the maximum range of values allowed for $l$: $\frac{q}{2}(1+\frac{1}{2\pi})\leqslant l\leqslant q$.
Again, this range is maximum for the most stable configuration. On the contrary,
when $\beta=0$ we can only have $l=q$, and therefore it is not possible to
 reduce the number of quarks. For this value of the magnetic flux the
strings are no longer bounded and the configurations reduce to $q$ free quarks.
Indeed, $\beta=0$, $l=q$ implies $\rho_1=\rho_0\to\rho^\prime=\infty$,
i.e. the fundamental strings become radial. Note as well that when $l=l_{\rm min}$
the strings become radial for any value of the magnetic flux.
The conclusion is that the $(l,{\cal N})$ parameter space for which the classical
configurations exist is bounded by those values corresponding to the free quarks case.

Equations (\ref{eqmotion3}) and (\ref{boundary2}) allow to calculate the radius of the configuration,
\begin{equation}
\label{size}
{\ell}=\frac{L^2\rho_1^2}{12\rho_0^3}\
\int_1^\infty \frac{dz}{z^2\sqrt{z^4-\frac{\rho_1^4}{\rho_0^4}}}=
\frac{L^2\rho_1^2}{12\rho_0^3}\
 {_2F_1}\left(\frac{1}{2},\frac{3}{4},\frac{7}{4};\frac{\rho_1^4}{\rho_0^4}\right)\ , 
\quad \frac{\rho_1^4}{\rho_0^4}=4\frac{l_{\rm min}}{l}\left(1-\frac{l_{\rm min}}{l}\right)\ ,
\end{equation}
where we have changed the integration variable as follows
$z=\frac{\rho}{\rho_0}$ and ${_2F_1}(a,b,c;x)$ is a hypergeometric function.
This expression has the same form than the size of the baryon vertex in
 $AdS_5 \times S^5$ \cite{BISY,JLR} and the $q{\bar q}$ system \cite{Maldacena,RY}.
Note that the dependence on the location of the D$p$-brane, $\rho_0$,
 and on $L^2$ is also the same.
 This is a non-trivial prediction of the AdS/CFT correspondence
 for the strongly coupled CS-matter theory. Note as well that (\ref{size})
 reduces to the expression found in \cite{GLR} when $l=q$.

The total on-shell energy is in turn given by
\begin{eqnarray}
E&=& E_{Dp} + E_{lF1}+E_{(q-l)F1} = \nonumber\\
&=&l\, T_{F_1}\rho_0\Big(\frac{q}{l}
\sqrt{1-\beta^2}+\int_1^{\infty}dz\frac{z^2}
{\sqrt{z^4-\frac{\rho_1^4}{\rho_0^4}}}+\frac{q-l}{l}\int_0^1 dz \Big)\ .
\end{eqnarray}
The binding energy can then be obtained by subtracting the (divergent)
 energy of the constituents. Note that, as we have discussed before,
the free quarks configuration is degenerate, since it can be reached in
three cases: when the D$p$-brane is located at $\rho_0=0$
 (at this location the energy of the D$p$ vanishes), as in \cite{BISY},
 when $\beta=0$ ($\Leftrightarrow l=q$) and $\rho_0$ is arbitrary,
and when $l=l_{\rm min}$, for any $\beta$ and any $\rho_0$.
In all these cases the constituents contribute with an energy $l\, T_{F1}\int_0^{\infty}d\rho$
and the binding energy is given by:
\begin{equation}
\label{Ebinding}
E_{bin}=l\ T_{F1}\, \rho_0\Bigl\{-\, {_2F_1}\left(-\frac{1}{4},\frac{1}{2},\frac{3}{4};
 4\frac{l_{\rm min}}{l}\left(1-\frac{l_{\rm min}}{l}\right)\right)
+2 \frac{l_{\rm min}}{l}-1\Bigr\}\ .
\end{equation}
This expression has again the same form than the corresponding expressions in
 \cite{BISY,JLR,Maldacena,RY,GLR}\footnote{In this case we have added the
on-shell energy of the D$p$-brane.}. 
 Setting $x=l_{\rm min}/l$ 
 the configurations are maximally stable when $x$ is minimum, i.e. when $l=q$ and $\beta$
reaches its maximum value. This happens for zero magnetic flux for the D2 and D6-branes,
and for $\frac{{\cal N}}{L^2}=\frac{1}{2\pi}$ for the D4.


\no


From (\ref{size}) and (\ref{Ebinding}) we have that for all $l$ and ${\cal N}$
the binding energy of the baryon reads
\begin{equation}
E_{bin}=-f(x)\frac{(g_s N)^{2/5}}{{\ell}}\leqslant0
\end{equation}
since $f(x)\geqslant 0$. The behavior of $f(x)$ is depicted in Figure 2. Moreover the
binding energy satisfies the concavity condition $\displaystyle{\frac{dE}{dL}\geqslant0,\frac{d^2E}{dL^2}\leqslant0}$. Therefore
the force is manifestly attractive and increasing in magnitude. Note however that this was not 
necessarily expected for baryons, since in this case there is no analogue of the concavity condition 
for heavy quark-antiquark pairs \cite{con1,con2}.
The $1/{\ell}$ behavior is that dictated by conformal invariance,
whereas the non-analytical dependence on the 't Hooft
coupling $\lambda$ is the one predicted in \cite{DPY,CW,RSY,Suyama,MP,KWY,DT},
which hints at a universal behavior based on the conformal symmetry of the gauge theory.


\begin{figure}[!t]
\begin{center}
\begin{tabular}{cc}
\includegraphics[scale=.8]{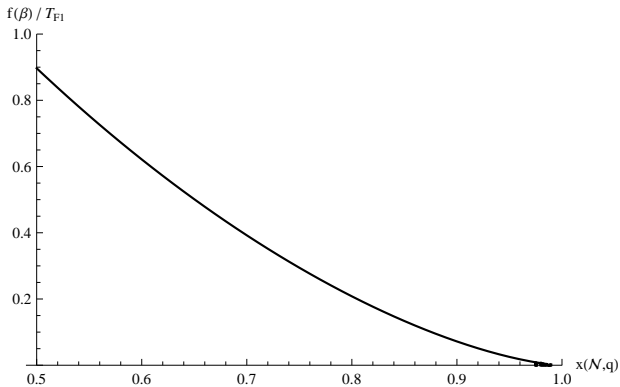}
\end{tabular}
\end{center}
\caption{Positivity of $f(x)$ as a function of $x$}
\label{fig:fbeta}
\end{figure}


\section{Stability analysis}

We shall next consider the stability analysis of the classical solution.
We know from \cite{SS} that the instabilities can emerge only from
 longitudinal fluctuations of the $l$ strings, since
only these possess a non-divergent zero mode, which is a sign of instability.
To study the fluctuations about the classical solution we perturb the embedding according to
\begin{equation}
r=r_{\rm cl}+\delta r(\rho)
\end{equation}
and expand the Nambu-Goto action to quadratic order in the fluctuations.
$\delta r$ is then solved from the equation
\begin{equation}
\frac{d}{d\rho}\Bigl(\frac{(\rho^4-\rho_1^4)^{3/2}}{\rho^2}\frac{d}{d\rho}\Bigr)\delta r=0\, ,
\end{equation}
from where we find
\begin{equation}
\delta r=A\int_{\rho}^\infty d\rho\frac{\rho^2}{(\rho^4-\rho_1^4)^{3/2}}
={\frac{A}{3\rho^3}}\ {_2F_1}\Bigl(\frac32,\frac34;\frac74;\frac{\rho_1^4}{\rho^4}\Bigr)\, .
\end{equation}
Supplementing with the boundary condition (eq. (3.12) in \cite{SS})
\begin{equation}
\rho_0 \gamma^2 \delta r^\prime+2 (1+\gamma^2)\delta r=0 \quad {\rm at} \quad \rho=\rho_0\ \quad
{\rm where} \quad \g\equiv\sqrt{1-\frac{\rho_1^4}{\rho_0^4}}\ ,
\end{equation}
we find that
\begin{equation}
_2F_1\Bigl(\frac32,\frac34;\frac74;1-\gamma^2\Bigr)=\frac{3}{2\gamma (1+\gamma^2)}\, .
\end{equation}
The numerical result for $\gamma$ is then $\gamma_c=0.538$.
The critical value for $a$ can be read from (\ref{boundary2}),
and we find it is a function of the magnetic flux
\begin{equation}
a_c=\frac{1+\sqrt{1-\beta^2}}{1+\gamma_c}
\end{equation}

Therefore, for the various configurations with magnetic flux there is a
 bound for the number of F-strings coming from stability
\begin{equation}
\label{stabilitybound}
l\geqslant \frac{q}{1+\gamma_c}(1+\sqrt{1-\beta^2})
\end{equation}
which is more restrictive than the bound imposed by the existence of a classical solution
\begin{equation}
l\geqslant \frac{q}{2}(1+\sqrt{1-\beta^2})\, .
\end{equation}
Note that in fact the stability condition (\ref{stabilitybound}) imposes
 a bound on the magnetic flux $1+\sqrt{1-\beta^2}\leqslant 1+\gamma_c$  which
 is also more restrictive than the one coming from (\ref{bound}), since
now $\beta\geqslant\sqrt{1-\gamma^2_c}$ and therefore $\beta=0$, which
was setting the condition for the maximum (and minimum, if applicable)
 magnetic flux, is not reached. Therefore stability further restricts
the allowed values for the magnetic flux coming from the analysis of the equations of motion.

Finally, we turn to the fluctuations of the $Dp$-brane. We
perturb the embedding according to
\ba
x^\m=\d x^\m(t,\th_\a)\ , \qq \rho=\rho_0\ ,\qq x^\m = x,y\ ,
\ea
leaving the position of the $Dp$-brane at $\rho=\rho_0$ intact due to the gauge choice $\rho=\s$ for the
strings. To be more precise, the $\rho$-fluctuations can be proven to be decoupled from the others both in the 
equations of motion and in the boundary equations; being periodic in the angles of $CP^{\frac{p}{2}}$.
Moreover, leaving the position of the brane at $\rho=\rho_0$ can also be proven to be allowed for spaces for which
$g_{tt}\sim \rho^2$ (as in $AdS_4$) at zero mode of the angular fluctuations, whereas for higher modes
the $\d\rho$ fluctuations are stable. Moreover, for the $CP^1$ and $CP^2$ cases we have kept fixed the D2 and D4 embeddings on the $CP^3$. We then find that
to second order in the fluctuations the expansion of the $Dp$-brane action reads
\ba
S_{Dp}&=&-\frac{T_p}{g_s} L^p (1+c)^{{p\ov4}}\int dt\,d\Om_p\sqrt{-g_{tt}}\sqrt{\g}
\bigg\{1+
{g_{\m\n}\ov 2(1+c)}\g^{\a\b}\del_\a\d x^\m\del_\b\d x^\n\nonumber\\
&&+{g_{\m\n}\ov 2g_{tt}}\d\dot{x}^\m\d\dot{x}^\n\bigg\}\ , \qq c=(2\pi\cN)^2\ ,
\ea
where $c=(2\pi({\cal N}-1))^2$ for the D2 and D6-branes, $c=(2\pi{\cal N})^2$ for the D4,
$\g_{\a\b}$ is the metric of $CP^{{p\ov2}}$ and the action is calculated at $\rho=\rho_0$. The
subscripts $\a,\m$ refer to the angles of $CP^{{p\ov2}}$ and to the $x,y$ coordinates, respectively.
Expanding the fluctuations in terms of the spherical harmonics of the $CP^{\frac{p}{2}}$ coset manifold
\footnote{Satisfying the eigenvalue equation $\nabla_\g^2\Psi_\ell=-\om^2_\ell\Psi_\ell$ where $\om_\ell^2$ 
is positive since the Laplace operator is defined on a compact manifold.} as
\be
\d x^\m(t,\th_\a) = \d x^\m(t) \Psi_\ell(\th_\a)\ ,
\ee
we find from the Euler--Lagrange equations for the action that
\ba
\label{1-7}
{d^2 \d x^\mu\ov dt^2}+\Om_\ell^2\d x^\mu=0, \qq \Om^2_\ell=-{g_{tt}\ov 1+c}\om^2_\ell\geqslant 0\ .
\ea
Note that there are no
boundary conditions for these fluctuations, the reason being that the $\mathbb{R}\times CP^{{p\ov2}}$ space
has no boundary. The conclusion is that the D$p$-brane is also stable against fluctuations.

\section{The microscopical description}

In the previous sections we have described magnetically charged baryon
vertex like configurations with varying number of quarks using the probe brane
approximation. This description is valid in the supergravity limit $L\gg 1$ (in string units),
equivalently when $k\ll N$, and in the
weakly coupled region in which $g_s\ll 1$, equivalently when $N\ll k^5$.
In this section we show that it is possible to give a description for finite 't Hooft coupling
in terms of fuzzy $CP^{\frac{p}{2}}$ manifolds built up out of dielectrically expanded D0-branes.

The fact that the magnetic flux induces D0-brane charge on the D$p$-branes
wrapped on $CP^{\frac{p}{2}}$ suggests a close analogy with the dielectric effect
 of \cite{Emparan,Myers}. We then expect that a complementary description
in terms of coincident D0-branes expanded into fuzzy $CP^{\frac{p}{2}}$ manifolds should be possible. 
This would be the `microscopical' realization of the `macroscopical' D$p$-branes
 wrapping classical $CP^{\frac{p}{2}}$ spaces with magnetic flux.
It is well known that the macroscopical and microscopical descriptions have
complementary ranges of validity \cite{Myers}. The first is valid in the
supergravity limit $L\gg 1$, whereas the second is a good description
when the mutual separation of the expanding D0-branes is much smaller than
the string length. For $n$ expanding such branes
this is fixed by the condition $L\ll n^{\frac{1}{p}}$. The two descriptions are
then complementary for finite $n$ and should agree in the large $n$ limit,
where they have a common range of validity.
In $AdS_4\times CP^3$ the regime of validity of the microscopical
description is fixed by the condition that $N\ll n^{\frac{4}{p}}\,k$.
Therefore this description allows to explore the region of finite 't Hooft coupling.

Dielectric branes expanding into fuzzy coset manifolds have been discussed in the literature in different contexts \cite{TV,JLR2,LR,JLR,JLR3}. $G/H$ coset manifolds can be described as fuzzy surfaces if $H$ is the isotropy group of the lowest weight state of a given irreducible representation of $G$ \cite{Madore,TV}. Since different irreducible representations have associated different isotropy subgroups they can give rise to different cosets $G/H$. For instance, $CP^2$ has $G=SU(3)$, $H=U(2)$, and this is precisely the isotropy group of the $SU(3)$ irreducible representations $(m,0)$, $(0,m)$, where we parameterize the irreducible representations of $SU(3)$ by two integers $(n,m)$ corresponding to the number of fundamental and anti-fundamental indices. Any other choice of $(n,m)$ has isotropy group $U(1)\times U(1)$ and therefore yields a different coset, $SU(3)/(U(1)\times U(1))$. One can also take a more geometrical view more suitable for our purposes. Using the fact that $CP^{\frac{p}{2}}$ spaces can be defined as the submanifolds of $\mathbb{R}^{\frac{p^2}{4}+p}$ determined by a given set of $p^2/4$ constraints, a fuzzy version arises by promoting the Cartesian coordinates that embed the $CP^{\frac{p}{2}}$ in $\mathbb{R}^{\frac{p^2}{4}+p}$ to $SU(\frac{p}{2}+1)$ matrices in the irreducible totally symmetric representations $(m,0)$ or $(0,m)$. Indeed only for these representations can the set of $p^2/4$ constraints be realized at the level of matrices.
The Cartesian coordinates are then taken to play the role of the non-Abelian transverse scalars that couple in Myers action for coincident D-branes. Using this action one can then provide a microscopical description of a D$q$-brane wrapped on the classical $CP^{\frac{p}{2}}$ space in terms of D$(q-p)$-branes expanding into a fuzzy $CP^{\frac{p}{2}}$. Exact agreement between the two descriptions is found in the large $m$ limit.

\subsection{The DBI action in the microscopical description}
\no
The DBI action describing the dynamics of $n$ coincident D0-branes is given by \cite{Myers}
\begin{equation}
\label{D0myers}
S^{DBI}_{nD0}=-\int d\tau\, {\rm STr}\Bigl\{e^{-\phi} \sqrt{|{\rm det}
\Bigl(P[E_{\mu\nu}+E_{\mu i}(Q^{-1}-\delta)^i{}_j
E^{jk}E_{k\nu}]\Bigr){\rm det}Q|}\Bigr\}
\end{equation}
where $E=g+B_2$,
\begin{equation}
\label{Qmatrix}
Q^i{}_j=\delta^i{}_j+\frac{i}{2\pi}[X^i,X^k]E_{kj}\, ,
\end{equation}
and we have set the tension of the D0-branes to 1.
We take $g_{\mu\nu}$ to be the metric in $AdS_4\times CP^3$ and $B_2=-2\pi J$, as in Appendix A.
The number of D0-branes, $n$, is related to the magnetic flux of
the macroscopical description by (\ref{nDqs}), with $s=0$
\be
\label{ene}
n=\frac{{\cal N}^{\frac{p}{2}}}{2^{\frac{p}{2}}(\frac{p}{2})!}\ .
\ee

 We now let these
 D0-branes expand into a fuzzy $CP^{\frac{p}{2}}$ space
 to build up a D$p$-brane.
We find that
\be
\label{DBID0}
S^{DBI}_{nD0}  =
 - \frac{1}{ g_s}\int d\tau {2\r\ov L}\ {\rm STr}\sqrt{\det(Q)} \ .
\ee

As we have mentioned, a fuzzy version of $CP^{\frac{p}{2}}$ is well-known. Here we will mainly follow \cite{CSW}.
 $CP^{\frac{p}{2}}$ is the coset manifold $SU(\frac{p}{2}+1)/U(\frac{p}{2})$,
and can be defined by the submanifold of $\mathbb{R}^{\frac{p^2}{4}+p}$
 determined by the set of $p^2/4$ constraints
\begin{equation}
\label{constraints}
\sum_{i=1}^{\frac{p^2}{4}+p}x^i x^i=1\, , \qquad \sum_{j,k=1}^{\frac{p^2}{4}+p} d^{ijk} x^j x^k=
\frac{\frac{p}{2}-1}{\sqrt{\frac{p}{4}(\frac{p}{2}+1)}}x^i
\end{equation}
where $d^{ijk}$ are the components of the totally symmetric $SU(\frac{p}{2}+1)$-invariant tensor.
The Fubini--Study metric of the $CP^{\frac{p}{2}}$ is given by
\begin{equation}
ds^2_{CP^{\frac{p}{2}}}=\frac{p}{4(\frac{p}{2}+1)}\sum_{i=1}^{\frac{p^2}{4}+p}(dx^i)^2\, .
\end{equation}
A fuzzy version of $CP^{\frac{p}{2}}$ can then be obtained by imposing the
 conditions (\ref{constraints}) at the level of matrices. This
is achieved with a set of coordinates $X^i$ ($i=1,\dots ,\frac{p^2}{4}+p$)
 in the irreducible totally symmetric representation of order $m$, $(m,0)$, satisfying
\begin{equation}
[X^i, X^j]=i \Lambda_{(m)}f_{ijk} X^k\, ,
\qquad \Lambda_{(m)}=\frac{1}{\sqrt{\frac{pm^2}{4(\frac{p}{2}+1)}+\frac{p}{4}m}}
\end{equation}
with $f_{ijk}$ the structure constants in the algebra of the generalized
Gell-Mann matrices of $SU(\frac{p}{2}+1)$.
The dimension of the $(m,0)$ representation is given by
\begin{equation}
\label{dimension}
\dim(m,0)=\frac{(m+\frac{p}{2})!}{m!(\frac{p}{2})!}\ .
\end{equation}
The K\"ahler form of the fuzzy $CP^{\frac{p}{2}}$ is given by (see Appendix B):
\begin{equation}
\label{fuzzykahler}
J_{ij}=\frac{1}{\frac{p}{2}+1}\sqrt{\frac{p}{4(\frac{p}{2}+1)}} f_{ijk} X^k\, .
\end{equation}

Substituting this non-commutative ansatz in (\ref{DBID0}) we can compute $\det(Q)$. 
This is however a difficult computation to perform in general, since
$Q^i{}_j=\delta^i{}_j+M^i{}_j$ with $M$ given by
\begin{equation}
\label{Mmatrix}
M^i{}_j=-\frac{1}{\frac{p}{2}+1}\Lambda_{(m)}f_{ikl}X^l\, 
\Bigl( \frac{pL^2}{8\pi}\delta^k{}_j-\sqrt{\frac{p}{4(\frac{p}{2}+1)}}f_{kjm} X^m\Bigr)\, ,
\end{equation}
and one has to compute traces
of powers of $M$ using the constraints above as well as (\ref{2-2}).
Given this we are going to start by making the comparison with the macroscopical calculation. 
For this purpose it is enough to work to leading order  in $m$, to which the second term in (\ref{Mmatrix}), 
coming from $B_2$, does not contribute. This should match the macroscopical result for $B_2=0$. Indeed, recall from section 2.1 that $B_2$ contributes to (\ref{energymac}) to order $O(1/{\cal N})$.
Already in this case we find that
\ba
\label{2-3}
&&{\rm Tr}(M)=0\ ,\qq {\rm Tr}(M^2)=-{p\ov 2^4\pi^2}\, r\ \mathbb{I}\ ,
\qq {\rm Tr}(M^3)=-i\,{p({p\ov2}+1)\ov2^7\pi^3L^2}\,r^2\ \mathbb{I}\ ,\\
&&{\rm Tr}(M^4)={p\ov 2^8\pi^4}\ r^2\ \mathbb{I}+
{p\ov 2^{10}\pi^4L^4}\left(\left({p\ov2}+1\right)^2-4\right)r^3\ \mathbb{I}\ ,
\nonumber
\ea
with
\begin{equation}
 r={L^4\ov m(m+{p\ov2}+1)}\, .
\label{lmfo}
 \end{equation}
However, in the limit
\be
L\gg 1\ ,\qq m\gg 1\ ,\quad {\rm with}\quad r\simeq {L^4\ov m^2}= {\rm finite}\ ,
\label{lmfi}
\ee
some terms in the traces of higher powers of $M$ drop out, and we find
\begin{equation}
{\rm Tr}(M^{2n})= p\, (-1)^n \left(\frac{r}{16\pi^2}\right)^n  \,\mathbb{I} \, ,
\qquad {\rm Tr}(M^{2n+1})=0 \ .
\end{equation}
Substituting in (\ref{DBID0}) we then obtain that
\begin{equation}
\det (Q)= \Bigl(1+\frac{r}{16\pi^2}\Bigr)^{\frac{p}{2}} \mathbb{I}\ .
\end{equation}

The DBI action of $n$ D0-branes expanding into a fuzzy $CP^{\frac{p}{2}}$ is then given to leading order in $m$ by
\begin{equation}
\label{DBImicro}
S^{DBI}_{nD0}=-\frac{n}{g_{s}}\Bigl(1+\frac{L^4}{16\pi^2 m^2}\Bigr)^{\frac{p}{4}}\int d\tau \frac{2\rho}{L}
\end{equation}
where $n=\dim(m,0)$ arises as $\dim(m,0)={\rm STr}\, \mathbb{I}$. Note that in the regime of
validity of the microscopical description $L\ll n^{\frac{1}{p}}\to L^4\ll m^2$, and we could expand in powers of 
$\displaystyle\frac{L^4}{m^2}$. 
We will see however that the agreement with the macroscopical description still holds for the entire 
expression in (\ref{DBImicro}). We encountered already this situation in the microscopical descriptions 
of giant gravitons in \cite{JL,JLR4,JLR2,JLR3}.
Taking into account (\ref{dimension}) and (\ref{ene}) we have that to leading order in $m$
the label of the irreducible representation and the unit of magnetic flux are related through
\begin{equation}
m\sim \frac{\cal N}{2}
\end{equation}
and (\ref{DBImicro}) becomes
\begin{equation}
S^{DBI}_{nD0}=-\frac{T_p}{g_{s}}\, {\rm Vol}(CP^{\frac{p}{2}})
\Bigl(L^4+(2\pi{\cal N})^2\Bigr)^{\frac{p}{4}}\int d\tau \frac{2\rho}{L}\ ,
\end{equation}
which exactly matches the result (\ref{energymac}) of the macroscopical calculation for $B_2=0$.
 Note that ${\cal N}\sim 2m$ is in agreement with the quantization condition ${\cal N}\in 2\mathbb{Z}$.

Let us now include the effect of the $B_2$ field. We know from the macroscopical calculation that
$B_2$ produces a shift ${\cal N}\rightarrow {\cal N}-1$ in the D2 and D6-branes, and cancels the contribution
of the Freed-Witten worldvolume flux in the D4-brane. Its effect is therefore $O(1/m)$, and this is why we could 
ignore it in the leading order calculation above. Analytical and numerical results for $B_2\ne 0$  and the agreement with the macroscopical calculation suggest
that the complete expression for the determinant to order $O(1/m)$ can be obtained from the expansion of
\begin{equation}
\label{detQB2}
{\rm det}(Q)=\left(\left(1-\frac{1}{2\sqrt{m(m+\frac{p}{2}+1)}}\right)^2+\frac{r}{16\pi^2}\right)^{\frac{p}{2}}\, .
\end{equation}
This is the exact result for $p=2$ in the limit (\ref{lmfi}) and correctly matches the macroscopical result to this order for all $p$.
 Indeed, using (\ref{detQB2}) we find that
\begin{equation}
\label{DBImicroB2}
S^{DBI}_{nD0}=-\frac{n}{g_{s}}\left(\left(1-\frac{1}{2\sqrt{m(m+\frac{p}{2}+1)}}\right)^2+\frac{L^4}{16\pi^2 m(m+\frac{p}{2}+1)}\right)^{\frac{p}{4}}
\int d\tau \frac{2\rho}{L}\, ,
\end{equation}
which to order $O(1/m)$ yields
\begin{equation}
S^{DBI}_{nD0}=-\frac{T_p}{g_{s}}\, {\rm Vol}(CP^{\frac{p}{2}})
\Bigl(L^4+(2\pi)^2 (2m+\frac{p}{2}+1-1)^2\Bigr)^{\frac{p}{4}}\int d\tau \frac{2\rho}{L}\ .
\end{equation}
Here we have not cancelled the two ones inside the parenthesis to emphasize their different origin, coming from the $1/m$ expansion of the second term in (\ref{DBImicroB2}) (the +1) and the $B_2$ contribution (the -1).
Comparing to the macroscopical calculation for $B_2\ne 0$ this result suggests a redefinition of ${\cal N}={\cal N}(m)$ to order $O(1/m)$:
\begin{eqnarray}
\label{N1}
{\cal N}&=&2m+\frac{p}{2}+1 \qquad {\rm for}\quad p=2,6\\
\label{N2} {\cal N}&=&2m+\frac{p}{2} \qquad {\rm for}\quad p=4
\end{eqnarray}
With these redefinitions we can, on the one hand, obtain a magnetic flux properly quantized, i.e. such that ${\cal N}\in 2\mathbb{Z}$, and, on the other hand, reproduce the expected shift of ${\cal N}$, ${\cal N}\rightarrow {\cal N}-1$, for $p=2,6$. The $p=4$ case is more interesting. Recall that in the macroscopical analysis $B_2$ was introduced in order to cancel the flux of the (Freed-Witten) vector field required by the Freed-Witten anomaly, such that ${\cal F}=F_{FW}+\frac{1}{2\pi}B_2=0$ \footnote{In fact, the original argument supporting this $B_2$-field in \cite{AHHO} had to do with the analysis of the supergravity charges, while the analysis of the D4-brane worldvolume dynamics arose as a consistency check. We refer to the original paper for more details.}. Microscopically we should see, in the absence of $B_2$, an obstacle to the expansion of the D0-branes into a $CP^2$, which should be absent for the $CP^1$ and $CP^3$. However, since the Freed-Witten field strength cannot couple in the worldvolume of D0-branes it is not clear a priori how exactly a non-vanishing $B_2$ could allow the construction of the $CP^2$.  We have found through a simple classical computation that $B_2$ is required in order to get an even ${\cal N}$, that is later interpreted as (twice) the units of magnetic flux in the macroscopical description. This clarifies the precise way in which the flat half-integer $B_2$ allows for the correct construction of the di-baryon with magnetic charge at the microscopical level.
We will see in the next section that the analysis of the charges carried by the different branes confirms the redefinitions (\ref{N1}), (\ref{N2}).

In conclusion, we have seen that it is indeed possible to give a microscopical
 description of the magnetic baryon vertex like
configurations of \cite{GLR} in terms of D0-branes expanding into fuzzy $CP^{\frac{p}{2}}$.
This expansion is caused by the couplings in
 the Born-Infeld part of the action, and therefore it is entirely due to a
gravitational dielectric effect, analogous to the one described in \cite{DRG,JLR}.
The regime of validity is fixed by the condition
\begin{equation}
N\ll k\, \left[\frac{(m+\frac{p}{2})!}{m!(\frac{p}{2})!}\right]^{\frac{4}{p}}\ .
\end{equation}
Therefore for finite $m$ this description allows to explore the region of finite 't Hooft coupling. Note however that for $B_2\ne 0$ we have not been able to give exact analytical expressions beyond the constant term in a $1/m$ expansion.

\subsection{The F-strings in the microscopical description}

An essential part of the baryon vertex-like configurations described in
this paper are the fundamental strings that stretch from the D$p$-brane
 to the boundary of $AdS_4$. In this section we show how these strings arise in the microscopic setup.

The CS action for $n$ coincident D0-branes is given by
\begin{equation}
\label{CSaction}
S_{CS}=\int_{\mathbb{R}}{\rm STr}
\Bigl\{ P \Bigl(e^{\frac{i}{2\pi}(i_X i_X)}\, \sum_q C_q \,\, e^{B_2}\Bigr) e^{2\pi F} \Bigr\}\ .
\end{equation}
In this expression the dependence of the background potentials on the
 non-Abelian scalars occurs through the Taylor expansion
\cite{GM}
\begin{equation}
\label{Taylorex}
C_q(t,X)=C_q(t)+X^k \partial_k C_q(t)+\frac12 X^l X^k \partial_l \partial_k C_q (t)+\dots
\end{equation}
and it is implicit that the pull-backs into the worldline are taken with gauge covariant derivatives
$D_t X^\mu=\partial_t X^\mu+i[A_t, X^\mu]$.

In the $AdS_4 \times CP^3$ background we have
\begin{equation}
F_2=\frac{2L}{g_s}J\, , \qquad F_6=\frac{L^5}{g_s}J\wedge J\wedge J\, , \qquad B_2=-2\pi J
\end{equation}
with $J$ the K\"ahler form of the $CP^3$. Therefore
taking into account (\ref{Taylorex}) the relevant CS couplings in this background are
\begin{eqnarray}
\label{dieCS}
S_{CS}&=&i \int d\tau {\rm STr}\Bigl\{ \Bigl[(i_X i_X)F_{2} - \frac{1}{(2\pi)^2}(i_X i_X)^3 F_{6}
+\frac{i}{2\pi}(i_X i_X)^2 F_2\wedge B_2- \nonumber\\
&&-\frac{1}{2} \frac{1}{(2\pi)^2}(i_X i_X)^3 F_2\wedge B_2\wedge B_2\Bigr] A_\tau\Bigr\}\, .
\end{eqnarray}
These terms arise, respectively, from
\begin{equation}
S_{CS}=\int  {\rm STr} \Bigl\{ P \Bigl(C_1  -\frac12 \frac{1}{(2\pi)^2} (i_X i_X)^2 C_5
+\frac{i}{2\pi}(i_X i_X)C_1\wedge B_2-\frac14 \frac{1}{(2\pi)^2}(i_X i_X)^2 C_1\wedge B_2\wedge B_2\Bigr)\Bigr\}
\end{equation}
in (\ref{CSaction}).

The first coupling in (\ref{dieCS}) is non-vanishing when the D0-branes
 expand into a fuzzy $CP^1$, which can be that in which a D2-brane is wrapped or
any of the $CP^1$ cycles of a $CP^2$ D4-brane or a $CP^3$ D6-brane.
Since the K\"ahler form for a fuzzy $CP^{\frac{p}{2}}$ is given by (see the Appendix B)
\begin{equation}
J_{ij}=\frac{1}{\frac{p}{2}+1}\sqrt{\frac{p}{4(\frac{p}{2}+1)}} f_{ijk} X^k
\end{equation}
we find that
\begin{equation}
\label{CS1}
S_{CS_1}=i\int STr\{(i_X i_X)F_2\wedge A\}=k \Bigl(m(m+\frac{p}{2}+1)\Bigr)^{-1/2}
\frac{(m+\frac{p}{2})!}{m! (\frac{p}{2})!}\int d\tau A_\tau
\end{equation}
which gives in the large $m$ limit
\begin{equation}
\label{kchargemic}
S_{CS_1}=k\, \frac{m^{\frac{p}{2}-1}}{(\frac{p}{2})!}\int d\tau A_\tau
\end{equation}
Taking into account that the dimension of the irreducible representation
is related to the units of magnetic flux of the macroscopical description
by $m=\frac{{\cal N}}{2}$, as we showed in the previous section, we find
 that the number of fundamental string charge in each $CP^1$ is given by:
\begin{equation}
\label{Fstring1}
q=\frac{2}{p}\, k \, \frac{{\cal N}^{\frac{p}{2}-1}}{2^{\frac{p}{2}-1}(\frac{p}{2}-1)!}
\end{equation}
which is in agreement with the macroscopical result (\ref{kcharge}).

Let us now look at the second term in (\ref{dieCS}). This term is
 non-vanishing when the D0-branes expand into a fuzzy $CP^3$, so
 it should give the fundamental string charge carried by the $CP^3$
D6-brane in the large $m$ limit. The explicit computation gives
\begin{equation}
\label{CS2}
S_{CS_2}=-\frac{i}{(2\pi)^2}\int  STr\{(i_X i_X)^3 F_{6}\wedge A\}=
N\Bigl(m(m+4)\Bigr)^{-3/2}\frac{(m+3)!}{m!}\int d\tau A_\tau
\end{equation}
and, in the large $m$ limit
\begin{equation}
\label{Ncharge}
S_{CS_2}=N \int d\tau A_\tau\, ,
\end{equation}
in agreement with the macroscopical result.

The third and fourth terms in (\ref{dieCS}) contribute when we take into
account the $B_2$ field that is necessary to compensate the Freed-Witten
worldvolume field of the D4-brane. Therefore they contribute to the $k$ charge to order $O(1/m)$ relative to 
(\ref{kchargemic}).
We find, explicitly:
\begin{equation}
\label{CS3}
S_{CS_3}= -\frac{1}{2\pi}\int  {\rm STr}\Bigl\{ (i_X i_X)^2 F_2\wedge B_2\wedge A\Bigr\}=
-k\Bigl(m(m+\frac{p}{2}+1)\Bigr)^{-1}\frac{(m+\frac{p}{2})!}{m!\, (\frac{p}{2})!}\int d\tau A_\tau
\end{equation}
and
\begin{eqnarray}
\label{CS4a}
S_{CS_4}&=&-\frac{i}{2}\frac{1}{(2\pi)^2}\int  {\rm STr}
\Bigl\{ (i_X i_X)^3 F_2\wedge B_2\wedge B_2\wedge A\Bigr\}=\nonumber\\
&=&\frac{3!}{8}k\Bigl(m(m+\frac{p}{2}+1)\Bigr)^{-3/2}\frac{(m+\frac{p}{2})!}
{m!\,(\frac{p}{2})!}\int d\tau A_\tau
\end{eqnarray}
These yield in the large $m$ limit
\begin{equation}
S_{CS_3}=-k\, \frac{m^{\frac{p}{2}-2}}{(\frac{p}{2})!}\int d\tau A_\tau
\end{equation}
and
\begin{equation}
\label{CS4}
S_{CS_4}=\frac{3!}{8}\, k \, \frac{m^{\frac{p}{2}-3}}{(\frac{p}{2})!}\int d\tau A_\tau
\end{equation}
respectively. In order to find the total $k$ charge to this (lower) order in $m$ (relative to (\ref{kchargemic})) we 
have to add the contributions to this order coming from  (\ref{CS1}), that we have ignored in (\ref{kchargemic}). Doing this 
we find that the total F-string charge for $p=2$ is still $k$, but for $p=4$ and $p=6$ it is given by  $k(m+1)$, 
$N+\frac{k}{2}((m+2)^2-m-2+\frac14)$, respectively. Taking into account the redefinitions (\ref{N1}) and (\ref{N2}) we find 
precisely the $k{\cal N}/2$ units of F-string charge of the $CP^2$ D4-brane and the $N+k\frac{{\cal N}({\cal N}-2)}{8}$ units 
of F-string charge of the $CP^3$ D6-brane, given respectively by equations (\ref{kcharge}) (for $p=4$) and (\ref{qD6}). Note 
that we find in addition a $k/8$ contribution for the D6, coming from $S_{CS_4}$. Macroscopically we already 
encountered this charge when computing the contribution of the coupling $\int_{D6} F_2\wedge B_2\wedge B_2\wedge A$ to the D6-brane tadpole. Given that
this charge was cancelled from the anomalous higher curvature coupling
\begin{equation}
S_{h.c.}=\frac32 (2\pi)^5\, T_6 \int d^7\xi \, P \Bigl(C_1\wedge
\sqrt{{\hat{A}(T)\ov\hat{A}(N)}}\ \Big) \wedge F\ ,
\end{equation}
a similar cancellation should occur microscopically. We will discuss in the next section how this can be achieved. 
Coming back to the D4-brane it is interesting that we need again at the classical level a flat half-integer $B_2$ in
 order to recover the right  fundamental string charge of the macroscopic D4-brane.

\subsection{Dielectric higher-curvature terms}

In this section we show that generalizing the microscopical Chern-Simons action in \cite{Myers} to include higher curvature terms 
\cite{GHM,CY,BBG} we can predict the existence of a dielectric higher curvature coupling in the action for multiple D0-branes that
 exactly cancels the $k/8$ contribution
to the D6-brane tadpole that we obtained above.

Generalizing the Chern--Simons action for multiple D$p$-branes in \cite{Myers} to include higher curvature terms we find
\ba
\label{hcdielectric}
S_{h.c.}=T_p \int d^{p+1}\xi \ S\tr\left[P\left(e^{\frac{i}{2\pi} (i_X i_X)}\sum_q C_q\ e^{B_2}\ \Omega\right)e^{2\pi F}\right]_{p+1}\, ,
\ \Omega=\sqrt{{\hat{A}(T)\ov\hat{A}(N)}}\ .
\ea
Keeping the first term in the $\hat{A}$-roof (Dirac) genus expansion, a general 
term of the previous expression for $D0$-branes has the following form
\ba
&&[(i_Xi_X)^nC_q\ (B_2)^k \Om_4]\wedge F^{\ell}\ ,\quad (n,\ell,k)\in\mathbb{N}\ , \\
&&\underbrace{(q+2(k-n)+4)}_{\geqslant0}+2\ell=1\ , \nonumber
\ea
where $\Om_4$ is given in term of the Pontryagin classes of the normal and the tangent bundle of the
three $CP^2$ circles of the $CP^3$ manifold \cite{ Eguchi,Bergman:2009zh}; $\Om_4=3(1-3){(2\pi)^4\ov48\pi^2}J\wedge J$.
To find the term of the expansion that contributes for the $CP^3$ we proceed as follows: We first note that
$\ell=0$ and that in the macroscopic limit only terms with $n+1=3\to n=2$ contribute, thus we have to solve $q+2k=1$,
which has solution $(k,q)=(0,1)$. Thus the term reads
\be
\label{anomalous}
S_{h.c.}=-{1\ov 2(2\pi)^2}\int_{\mathbb{R}} P[(i_Xi_X)^2C_1\wedge \Om_4]=
-{i \ov (2\pi)^2}\int_{\mathbb{R}} [(i_Xi_X)^3 (F_2\wedge\Om_4)] A
\ee
and substituting $F_2$ and $\Omega_4$:
\be
S_{h.c.}=-{\k\ov8}
(m(m+4))^{-3/2}{(m+3)!\ov m!}\int_{\mathbb{R}} d\tau A_\tau\simeq -{\k\ov8}\ \int_{\mathbb{R}} d\tau A_\tau\ ,
\ee
where we took into account that there are three $CP^2$ circles in $CP^3$. Thus this higher curvature coupling
cancels the $S_{CS_4}$ contribution as in the macroscopical case.

Anomalous dielectric couplings as those predicted by (\ref{hcdielectric}) have, to the best of our
 knowledge, not been discussed before in the literature. Furthermore, acting with T-duality on the A-roof in (\ref{hcdielectric}) one can obtain dielectric terms that couple 
the RR-potentials to derivatives of $B_2$ and the metric that generalize the anomalous terms derived in \cite{Becker,Garousi} for a single 
D$p$-brane. It would be interesting to confirm the existence of all these new couplings through string amplitude calculations.

\subsection{Stability analysis}

The study of the stability goes along the same lines than in the macroscopical set-up. Note that also in the microscopical description the DBI action can be written as (\ref{energymac}), where $Q_p$ depends now on the label of the irreducible representation, $m$, in the precise way given by (\ref{DBImicroB2}). The number of F-strings that must end on the D$p$-brane is in turn given by 
the sum of the contributions from equations (\ref{CS1}), (\ref{CS2}), (\ref{CS3}) and (\ref{CS4a}), where some of these terms have to be multiplied by the number of $CP^1$ or $CP^2$ cycles in the $CP^3$ as appropriate. Other than these differences we can vary the number of quarks, study the dynamics and the stability exactly along the same lines as in sections 3 and 4. Only now equation (\ref{bound})  will impose a bound on $m$, that is, on the number of D0-branes that can expand into a fuzzy $CP^{\frac{p}{2}}$ by Myers dielectric effect. In the large $m$ limit this is the bound that we encountered for ${\cal N}$ in the macroscopical description. As in there the existence of this bound should be related in some way to the stringy exclusion principle of \cite{MS}, although we have not been able to find a direct interpretation. 

The conclusion is that also in the microscopical set-up there exist perfect baryon vertex classical solutions to the equations of motion that are stable against fluctuations.

\section{Conclusions}

We have analyzed various configurations of
 magnetically charged particle-like branes in ABJM with reduced number of quarks.
We have shown that 't Hooft monopole, di-baryon and baryon vertex
 configurations with magnetic charge and reduced number of quarks
 can be constructed which are not only perfect classical solutions
 to the equations of motion but also stable against small fluctuations.

The magnetic flux has to satisfy some upper bound (also some lower
bound for the di-baryon, consistently with the fact that the D4 with
fundamental strings only exists for non-zero magnetic flux), and once this bound
is fixed it is possible to reduce the number of quarks to a minimum
value determined by ${\cal N}$ (or $\beta$):
$$l\geqslant \frac{q}{2}(1+\sqrt{1-\beta^2})$$
From here we can see that the number of quarks is maximally reduced
 when the energy of the configuration is minimum, that is, for those values of
 the flux for which $\beta=0$.

The analysis of the stability against small fluctuations reveals
 that the configurations are stable if
$$l\geqslant \frac{q}{1+\gamma_c}(1+\sqrt{1-\beta^2})$$
where $\gamma_c$ is fixed numerically to $\gamma_c=0.538$. Stability
 therefore increases the classical lower bound for each value of the magnetic flux.
 This is the same effect encountered in \cite{SS} for asymptotically $AdS_5 \times S^5$ spaces. 
It is worth mentioning that in fact following  \cite{SS} it
is trivial to extend our analysis to 
asymptotically $AdS_4 \times CP^3$ backgrounds and non-zero temperature. 

The previous analysis is based on a probe brane approximation, and is
 therefore valid in the supergravity limit $k\ll N$. Using the fact that
we can consistently add dissolved D0-branes to the configurations we have
given an alternative description in terms of D0-branes expanded into
 fuzzy $CP^{\frac{p}{2}}$ spaces that allows to explore the finite 't Hooft coupling region. In this description the expansion is caused by a purely
gravitational dielectric effect, while the Chern-Simons terms only
indicate the need to introduce the number of fundamental strings
required to build up the (generalized) vertex. The microscopical
 analysis confirms the existence of non-singlet classical stable solutions for finite 't Hooft coupling.
 
An output of this analysis is the prediction of dielectric higher
curvature couplings that to the best of our knowledge have not
been considered before
in the literature. The particular explicit coupling in the action for multiple D0-branes that has come
 out in our analysis is necessary in order to obtain the right
 fundamental string charge of the baryon vertex. For the
rest of branes they are predicted by
T-duality. These couplings imply in turn new couplings of the RR-potentials
 to derivatives of $B_2$ and the metric, along the lines in \cite{Becker,Garousi}, with further implications for other branes via S and U dualities.
 It would be interesting to explore more closely these implications.
 
Finally, it would be interesting to extend the existence of non-singlet baryon vertex like configurations like the ones considered in this paper to theories with reduced supersymmetry, like the Klebanov--Strassler
backgrounds \cite{KS}, where the internal geometry is the $T^{1,1}$ conifold.

\subsection*{Acknowledgements}

We would like to thank N. Guti\'errez, D. Rodr\'{\i}guez-G\'omez, D.C. Thompson and P. Tziveloglou for useful discussions.
The work of Y.L. and M.P. has been partially supported by the
research grants MICINN-09-FPA2009-07122, MEC-DGI-CSD2007-00042 and
COF10-03. K. Siampos has been supported by the ITN programme
PITN-GA-2009-237920, the ERC Advanced Grant 226371, the IFCPAR
CEFIPRA programme 4104-2 and the ANR programme blanc NT09-573739.
 K.Siampos would also like to thank the
University of Patras for
hospitality and the Theory Unit of CERN for
hospitality and financial support where part of this work was done.

\appendix

\section{Review of the $AdS_4\times CP^3$ background}

In this appendix we give a short review of the $AdS_4\times CP^3$ background.
In our conventions the $AdS_4\times CP^3$ metric reads
\begin{eqnarray}
ds^2
 =  L^2\Bigl( {1\over 4} ds^2_{AdS_4}+ds^2_{\mathbb{CP}^3}\Bigr)\, ,
\label{a.1}
\end{eqnarray}
with $L$ the radius of curvature in string units
\begin{equation}
L=\Bigl(\frac{32\pi^2 N}{k}\Bigr)^{1/4}\
\end{equation}
and where we have normalized the two factors such that $R_{\m\n}=-3 g_{\m\n}$ and $8 g_{\a\b}$ for
$AdS_4$ and $CP^3$, respectively.
The explicit parameterization of $AdS_4$ we use in the main text is
\begin{eqnarray}
ds_{AdS_4}^2 = \frac{16\,\rho^2}{L^2}d\vec x^2+L^2\frac{d\rho^2}{\rho^2}\ , \quad d\vec x^2=-d\tau^2+dx_1^2+dx_2^2\ .
\label{a.3}
\end{eqnarray}
For the metric on $CP^3$ we use the parameterization in \cite{Pope1984,Warner}
\begin{eqnarray}
ds_{\mathbb{CP}^3}^2=&&d\mu^2+\sin^2\mu\,\Big[ d\alpha^2+\frac{1}{4}\sin^2\alpha\,
 \big(\cos^2\alpha\,(d\psi-\cos\theta\, d\phi)^2+d\theta^2+\sin^2\theta\, d\phi^2\big)\nonumber \\
&&+\frac{1}{4}\cos^2\mu\, \big(d\chi+\sin^2\alpha\, (d\psi-\cos\theta\, d\phi)\big)^2\Big]\ ,
\end{eqnarray}
where
\begin{equation}
0\leqslant \mu,\,\alpha\leqslant\frac{\pi}{2}\, ,\quad 0\leqslant \theta \leqslant \pi\,
 ,\quad 0\leqslant \phi\leqslant 2\pi\, ,\quad 0\leqslant \psi,\, \chi\leqslant 4\pi\ .
\end{equation}
Inside $CP^3$ there is a $CP^1$ for $\m=\a=\pi/2$ and fixed $\chi$ and $\psi$ and
also a $CP^2$ for fixed $\th$ and $\phi$.

\no
In these coordinates the connection in $ds^2_{S^7}=(d\tau + \mathcal{A})^2 + ds^2_{\mathbb{CP}^3}$ reads
\begin{equation}
\mathcal{A}=\frac{1}{2}\sin^2\mu\, \Big(d\chi+\sin^2\alpha\,\big(d\psi-\cos\theta\, d\phi)\Big)\, .
\end{equation}
\no
The K\"ahler form
\begin{equation}
J=\frac{1}{2}d\mathcal{A}\, ,
\end{equation}
is then normalized such that
\begin{equation}
\int_{CP^1}\, J=\pi\, ,\qquad \int_{CP^2}\, J\wedge J=\pi^2\, , \qquad \int_{CP^3}\, J\wedge J\wedge J=\pi^3\, .
\end{equation}
Therefore,
\begin{equation}
\frac{1}{6}\,J\wedge J\wedge J= d{\rm Vol}(\mathbb{P}^3) \qquad {\rm and}
\qquad {\rm Vol}(\mathbb{CP}^3)=\frac{\pi^3}{6}\ .
\end{equation}
The $AdS_4 \times CP^3$ background fluxes can then be written as
\begin{equation}
F_2=\frac{2L}{g_s}J\, ,\qquad F_4=\frac{3L^3}{8 g_s}\ d{\rm Vol}(AdS_4)\, ,
\qquad F_6=-(\star F_4) =\frac{6\, L^5}{g_s}\ d{\rm Vol}(\mathbb{P}^3)\ ,
\end{equation}
where $\displaystyle g_s=\frac{L}{k}$.
The flux integrals satisfy
\begin{equation}
\label{fluxes}
\int_{CP^3}\, F_6=32\, \pi^5\, N\, ,\qquad \int_{CP^1}\,F_2=2\pi\, k\, .
\end{equation}
The flat $B_2$-field that is needed to compensate for the
Freed--Witten worldvolume flux in the D4-brane is given by \cite{AHHO}
\begin{equation}
B_2=-2\pi J\ .
\end{equation}

\section{Computation of the K\"ahler form for fuzzy $CP^{\frac{p}{2}}$}

In this Appendix we compute the K\"ahler form for the fuzzy  $CP^{\frac{p}{2}}$ spaces considered in the paper.
The K\"ahler form is given in terms of the exterior derivative of the one form
$U(1)$ gauge field \cite{Kahler1,Kahler2}
\begin{eqnarray}
\label{2-24}
&&J=J_{(i)} X^i\ ,\qquad J_{(i)}=\ha\ dA_i\ , \qquad A_i=\sqrt{{p\ov {p\ov2}+1}}\ L_i , \\
&&J\equiv\ha J{}_{ij}L_i\wedge L_j\ ,\qq L_i=-i\, {\rm Tr}(t_ig^{-1}dg)\ ,\qq g\in SU\left({p\ov2}+1\right)\ ,
\nonumber
\end{eqnarray}
where $t_i$ are the generators of $SU({p\ov2}+1)$ in the adjoint representation, $(t_i)_{jk}=-i f_{ijk}$.
Using that ${\rm Tr}(t_it_j)=({p\ov2}+1)\d_{ij}\ ,{\rm Tr}(t_it_jt_k)=i\ {{p\ov2}+1\ov2}f_{ijk}$, which result
from the identities \cite{deAzcarraga:1997}
\ba
\label{2-2}
&&f_{ikm}f_{jkm}=N\d_{ij},\ f_{iaj}f_{jbk}f_{kci}=-{N\ov2}f_{abc},\
d_{iaj}d_{jbk}f_{kci}={N^2-4\ov2N}f_{abc}, \nonumber\\
&&f_{iaj}f_{jbk}f_{kcm}f_{mdi}=\d_{ab}\d_{cd}+\d_{ad}\d_{bc}+{N\ov4}
(d_{abe}d_{cde}+d_{ade}d_{bce}-d_{ace}d_{bde})\ ,
\ea
we compute the K\"ahler form as follows
\ba
\label{2-25}
&&J={i\ov2}\sqrt{{p\ov {p\ov2}+1}}\ {\rm Tr}(t_kg^{-1}dg\wedge g^{-1}dg)X^k\ , \qq i L_it_i=\left({p\ov2}+1\right)\ g^{-1}dg \nonumber \\
&&\Rightarrow J={i^3\ov2}\sqrt{{p\ov {p\ov2}+1}}\ {{\rm Tr}(t_it_jt_k)\ov({p\ov2}+1)^2}\ X^k\ L_i\wedge L_j\ , \nonumber \\
&&\Rightarrow\qq J_{ij}={1\ov {p\ov2}+1}\sqrt{p\ov4({p\ov2}+1)}f_{ijk}X^k
\ea
Then, for the $n$ $D0$-branes expanding into a fuzzy $CP^{{p\ov2}}$ we find that
\ba
\label{2-26}
&&(i_X i_X)J=X^jX^iJ_{ij}=-{i\ov2}\sqrt{p\ov4({p\ov2}+1)}\ \L_{(m)}\mathbb{I}\ , \\
&&(i_X i_X)^{{p\ov2}}\underbrace{J\wedge J\wedge\cdots\wedge J}_{{p\ov2}\ \rm terms}=
\left({p\ov2}\right)!\left(-{i\ov2}\sqrt{p\ov4({p\ov2}+1)}\ \L_{(m)}\right )^{{p\ov2}}\mathbb{I}\ ,\nonumber
\ea
so that the interior products are constant.

\end{document}